\documentclass[11pt,twoside]{article}
\usepackage{./asp2014}{}

\aspSuppressVolSlug
\resetcounters

\begin{document}

\title{Probing the Structure of the Accretion Region in a Sample of Magnetic Herbig Ae/Be Stars}

\author{M.A. Pogodin$^1$, J.A. Cahuasqui$^2$, N.A. Drake$^2,^3$, S. Hubrig$^4$, M. Sch\"oller$^5$, \\
M. Petr-Gotzens$^5$, G.A.P. Franco$^6$, D.F. Lopes$^3$, O.V. Kozlova$^7$, B. Wolff$^5$, \\
J.F. Gonzalez$^8$, T.A. Carroll$^4$, S. Mysore$^5$
\affil{$^1$ Central Astronomical Observatory at
Pulkovo of Russian Academy of Sciences, Pulkovskoye chaussee 65, 196140,
Saint-Petersburg, Russia \email{pogodin@gao.spb.ru}}
\affil{$^2$ Sobolev Astronomical Institute, St. Petersburg State University,
   Universitetski pr. 28, 198504, Saint-Petersburg, Russia}
\affil{$^3$ Observat\'orio Nacional/MCTI, Rua Jos\'e Cristino 77, CEP 20921-400,
      Rio de Janeiro, RJ, Brazil}
\affil{$^4$ Leibniz-Institute f\"ur Astrophysik Potsdam (AIP), An der Sternwarte 18,
      14482 Potsdam, Germany}
\affil{$^5$ European Southern Observatory, Karl-Schwarzschild-Str.~2, 85748 Garching, Germany}
\affil{$^6$ Departamento de F\'\i sica, ICEx, UFMG, Caixa Postal 702,
    30.123-970 Belo Horizonte, MG, Brazil}
\affil{$^7$ Crimean Astrophysical Observatory, 98409, Nauchny, Crimea, Russia}
\affil{$^8$ Instituto de Ciencias Astronomicas, de la Tierra y del Espacio (ICATE)
5400, San Juan, Argentina}}

\paperauthor{M.A. Pogodin}{pogodin@gao.spb.ru}{}{Central Astronomical Observatory at
Pulkovo of Russian Academy of Sciences}{}{Saint-Petersburg}{}{196140}{Russia}
\paperauthor{J.A. Cahuasqui}{}{}{Sobolev Astronomical Institute, St.Petersburg State
University}{}{Saint-Petersburg}{}{198504}{Russia}
\paperauthor{N.A. Drake}{}{}{Observat\'orio National/MCTI}{}{Rio de Janeiro}{}{CEP 20921-400}{Brazil}
\paperauthor{S. Hubrig}{}{}{Leibniz-Institute f\"ur Astrophysik Potsdam
(AIP)}{}{Potsdam}{}{14482}{Germany} \paperauthor{M. Scholler}{}{}{European Southern
Observatory}{}{Garching}{}{85748}{Germany}
\paperauthor{M. Petr-Gotzens}{}{}{European Southern Observatory}{}{Garching}{}{85748}{Germany}
\paperauthor{G.A.P. Franco}{}{}{UFMG}{Departamento de F\'\i sica}{Belo
Horizonte}{MG}{Caixa Postal 702}{Brazil}
\paperauthor{D.F. Lopez}{}{}{Observatorio
National/MCTI}{}{Rio de Janeiro}{RJ}{CEP 20921-400}{Brazil} \paperauthor{O.V.
Kozlova}{}{}{Crimean Astrophysical Observatory}{}{Nauchnyi}{Crimea}{}{Russia}
\paperauthor{B. Wolff}{}{}{Leibniz-Institute fur Astrophysik Potsdam
(AIP)}{}{Potsdam}{}{14482}{Germany} \paperauthor{J.F.Gonzalez}{}{}{Instituto de
Ciencias Astronomicas de la Tierra, y del Espacio (ICATE)}{}{San Juan}{}{5400}
{Argentina} \paperauthor{T.A. Carroll}{}{}{Leibniz-Institute f\"ur Astrophysik Potsdam
(AIP)}{}{Potsdam}{}{14482}{Germany}

\begin{abstract}
We present  the results of a study of the temporal behaviour of several diagnostic lines
formed in the region of the accretion-disk/star interaction in the three magnetic Herbig
Ae stars HD\,101412, HD\,104237, and HD\,190073. More than 100 spectra acquired with
the ISAAC, X-shooter, and CRIRES spectrographs installed at the VLT-8m telescope (ESO, Chile),
as well as at other observatories (OHP, Crimean AO) were %
analyzed.
The spectroscopic data were obtained in the He\,{\sc i} $\lambda$10830,
Pa$\gamma$ and He\,{\sc i} $\lambda$5876 lines. We found that the temporal behaviour of the
diagnostic lines in the spectra of all program stars can be widely explained by
a rotational modulation of the line profiles generated by a local accretion flow.
This result is in good agreement with the predictions of the magnetospheric accretion
model.
For the first time, the rotation period of HD\,104237 ($P_{\rm rot}$ =
$5.37\pm0.03$ days), as well as the inclination angle ($i = 21^\circ\pm4^\circ$) were
determined.
Additional analysis of the HARPSpol spectra of HD\,104237 and
HD\,190073, taken from the ESO archive, with the use of the SVD method shows that
the magnetic field structure of HD\,190073 is likely more complex than a simple
dipole and contains a circumstellar component. For the first time, the magnetic field
of the secondary component of the binary system HD\,104237 was also detected
($\left<B_{\rm z}\right> = 128\pm10$~G).
\end{abstract}

\section{Introduction}
Herbig Ae/Be stars (HAeBes) are pre-main sequence (PMS) objects of intermediate mass
approximately from 2 to 8\,$M_{\odot}$
\citep{herbig1960,finkenzeller1984,the1994}.
This corresponds to the range of spectral classes F2 -- B0.
They are surrounded by dust/gas accretion disks.
Remote cold dust reveals itself in the form of a far-IR excess, and numerous emission lines
originate in the circumstellar (CS) envelope. This envelope has a complex
spatial structure and contains an equatorial accretion disk and matter outflows in
the form of a stellar/disk wind at higher latitudes.

One of the important unresolved problems in the HAeBes is the character of the interaction between
the accretion disk and the central star.
For the lower-mass PMS objects,
the classical T\,Tauri stars (CTTS) with a similar structure of their CS envelopes and
strong magnetic fields of the order of kG,  the magnetospheric accretion (MA) model
is generally recognized.
According to this model, the accretion disk does not contact directly the stellar surface,
but is truncated by the stellar magnetic field at some distance from the star.
A part of the accreted material falls onto the
star near %
magnetic pole regions along the closed magnetic  field lines, another part
outflows away along the open force lines \citep{tout1994}.
 However, the HAeBes have weaker magnetic fields (several 100\,G).
Is the MA model also applicable to them?

If the magnetic axis does not coincide with the rotation axis, a rotational
modulation of the line profiles has to be observed with a period $P$ equal to the
rotational period $P_{\rm rot}$ of the star. We tried to find signatures of such
modulation in the spectra of our program stars.

\section{Objects of the program}%1

HD\,101412 is an early Ae star with an unusually large magnetic field ($\sim3$~kG),
which is more typical for CTTS than for HAeBes. A magnetic dipole model of the
object has been suggested by \citet{hubrig2011}. It ensures a good fitting of
observational data, the $\left<B_{\rm z}\right>$  sine-like phase dependency
constructed with $P_{\rm rot}$ = $42.078\pm0.017^{d}$ is presented in Fig.\,4 of
\citet{hubrig2011}.

The second object of our program, HD\,104237, is a well-known binary system
(A:pe\,+\,K) with a T\,Tauri star as a secondary. The orbital solution of the system
has been obtained in \citet{bohm2004}: $P_{\rm orb}$ = 19.859$^{\rm d}$, $e$ = 0.66.
The object is the first HAeBe star for which the magnetic field was measured with
$\left<B_{\rm z}\right>\approx50$\,G \citep{donati1997}. Further observations and
work by \citet{wade2007} did not confirm this result, but recently
\citet{hubrig2013}
testified the presence of a weak magnetic field ($\left<B_{\rm
z}\right>$\,=\,$63\pm15$\,G). We can expect that the magnetic field of the object is
variable.

The last program object, HD\,190073, is a peculiar Herbig A2~IVpe star
(\citealt{pogodin2005}, and references therein). Its magnetic field
($\left<B_{\rm z}\right>\approx$\,100\,G) has been measured by various authors
\citep{hubrig2006,hubrig2009,catala2007}. There are clear indications of its
variability.

\section{Observations}

Three instruments installed at the VLT-8m telescopes (ESO, Chile) were used for
spectroscopic observations of the program stars in the near-IR region.

Eight spectra of HD\,104237 (March - April 2013) and five spectra of HD\,190073 (August -
September 2013) were acquired with ISAAC ($R=11\,500$). 13 spectra of HD\,101412
(December 2013 - March 2014), 13 spectra of HD\,104237 (November 2013 - January
2014) and six spectra of HD\,190073 (March - September 2010) were obtained using
X-shooter ($R\sim11\,000$).
These spectra cover a wide spectral range from the near-UV to
the near-IR and also include the He\,{\sc i} $\lambda$5876 line.
Nine additional spectra of HD\,101412 (April 2011 - March 2013) were acquired with CRIRES ($R = 110\,000$).
The $S/N$ ratio of all spectra was between 200 and 400.

 Additionally, 51 spectra of HD\,190073 near the He\,{\sc i}
$\lambda$5876 line were collected from archives of several observatories (ESO, OHP,
Crimean AO), which were obtained between 1994 and 2013
\citep{pogodin2005,pogodin2012,kozlova2014}.

\section{Results of the observations}

\articlefiguretwo{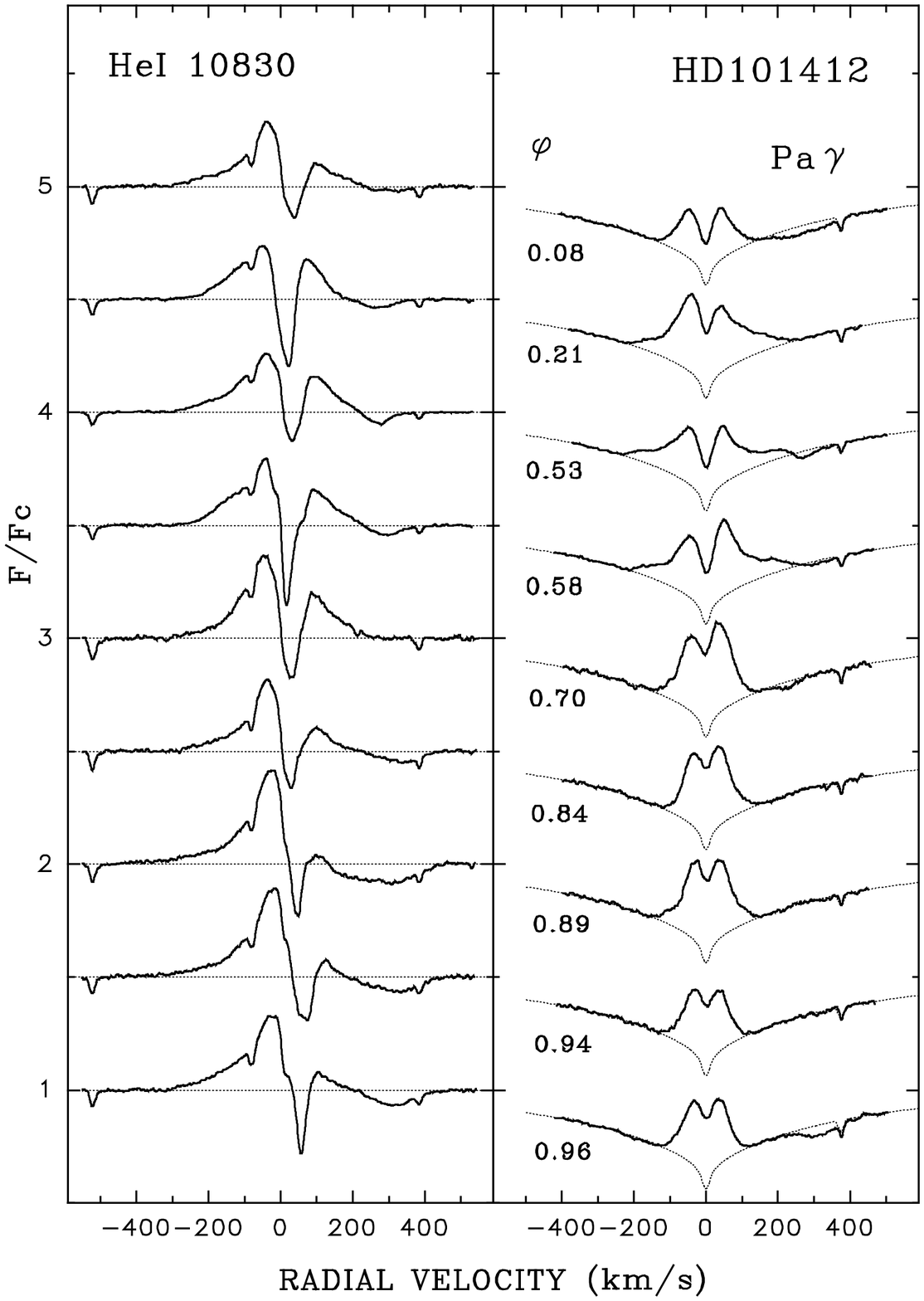}{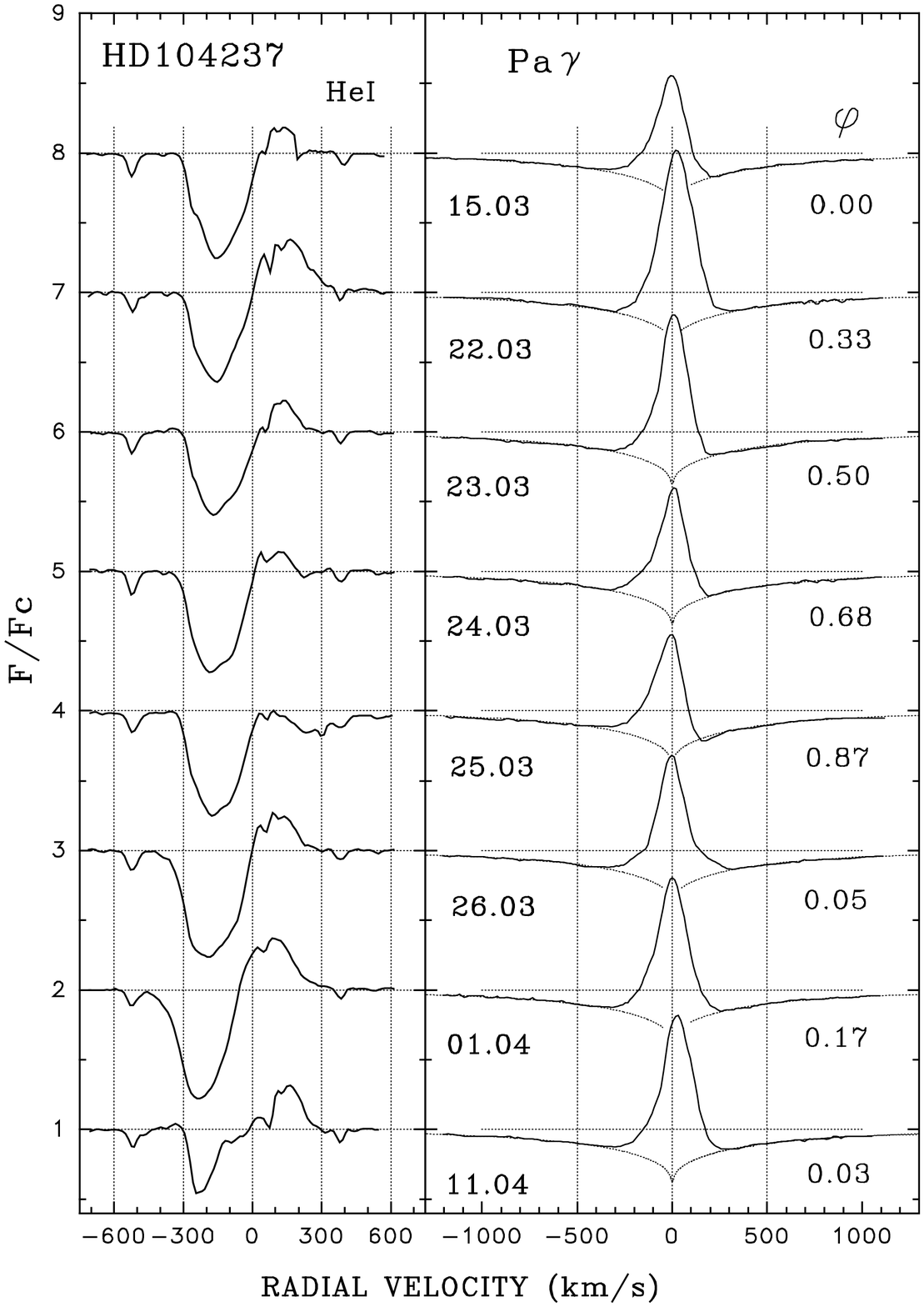}{ex-fig1}{\emph{Left:} Typical profiles of the
IR He\,{\sc i} $\lambda$10830 and Pa$\gamma$ lines in the spectrum of HD\,101412.
Phases of $P_{\rm rot}$ = $42.076^{\rm d}$ are
indicated. \emph{Right:} The same as in the left panel but for HD\,104237 and
$P_{\rm rot}$ = $5.37^{\rm d}$.}

Typical profiles of the two IR lines He\,{\sc i} $\lambda$10830 and Pa$\gamma$ in the
spectra of HD\,101412 (CRIRES) and HD\,104237 (ISAAC) are shown in
Fig.~\ref{ex-fig1}. The emission He\,{\sc i} line profile of HD\,101412 has two
redshifted absorption components.
One is wide, shallow,  and strongly redshifted. %
The second is narrow, deep  and with a weak red shift.
The emission profile of the Pa$\gamma$ line is a single one with a central
absorption.
The profiles of these two IR lines in the spectrum of HD\,104237
demonstrate quite different types. The He\,{\sc i} line profiles are of P\,Cyg type, which is
evidence for a strong stellar wind and an intermediate orientation of the rotation
axis relative to the line of sight.
The Pa$\gamma$ line does not show signs of wind.
Its density is not enough to be visible in this subordinate line.
As in the case of HD\,104237, HD\,190073 has similar IR line profiles: He\,{\sc i} $\lambda$10830
with a P\,Cyg structure and Pa$\gamma$ with a single emission profile. This object has
also an intermediate orientation relative to the observer. The He\,{\sc i} $\lambda$5876
line profiles appear as single emissions too.
\section{Analysis of the results}

 \articlefiguretwo{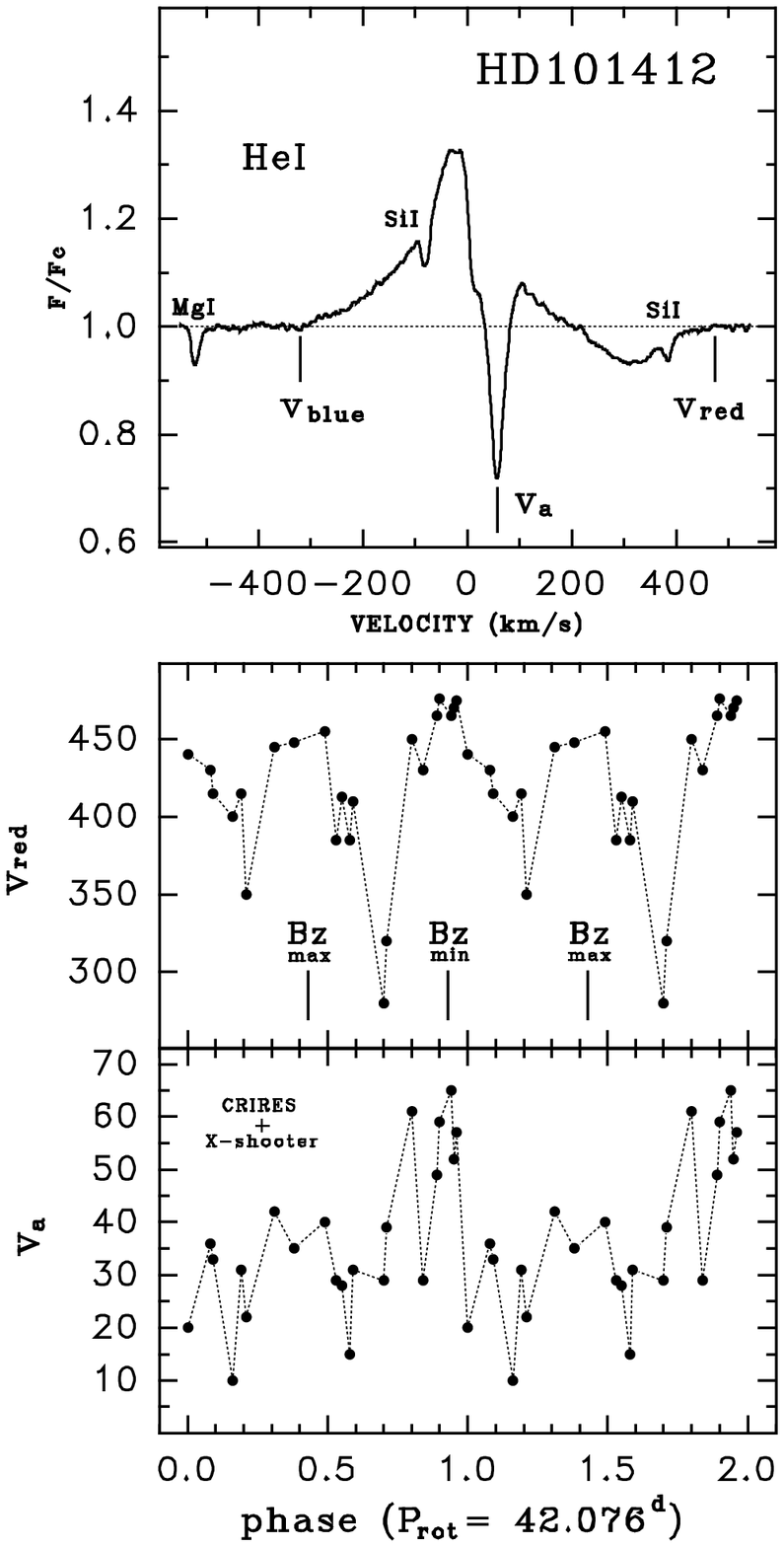}{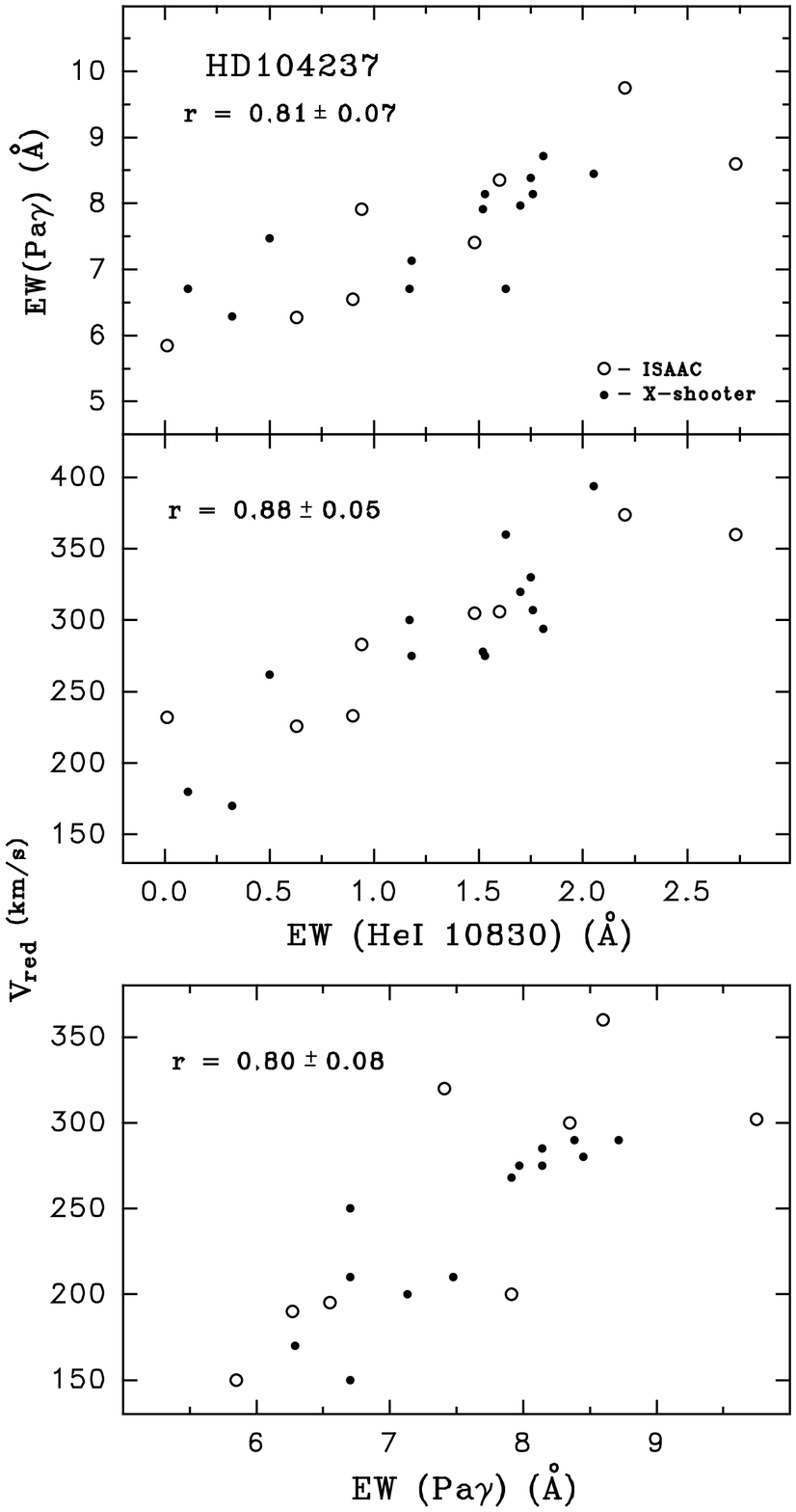}{ex-fig2}{\emph{Left:} Phase dependencies of
the velocities  $v_{\rm red}$ and $v_{\rm a}$ of the He\,{\sc i}  $\lambda$10830 profile in the
spectrum of HD\,101412. \emph{Right:} Dependencies between different parameters of the IR
lines in the spectrum of HD\,104237.
$EW$ of the Pa$\gamma$ line and $v_{\rm red}$ velocity versus $EW$ of the He\,{\sc i} 10830 line (top and middle panels respectively); $v_{\rm red}$ versus $EW$ of the Pa$\gamma$ line (bottom panel).
In the measurements of $EW$ of the  emission component of the He\,{\sc i} $\lambda$10830 line, we considered  emission above the underlying continuum, whereas in the case of the Pa$\gamma$ line, the
$EW$ of the emission component was measured as an emission above the preliminary
calculated photospheric Pa$\gamma$ absorption profile.
}

The model of the magnetic dipole-type field of HD\,101412 \citep{hubrig2011}
suggests that the angle between the rotation axis and the line of sight is
$i\,=\,80^\circ\pm7^\circ$, and the angle between the rotation and magnetic axes
is $\beta\,=\,84^\circ\pm13^\circ$.
With such an orientation, the magnetic axis has to be  located
close to the disk plane. The MA model predicts two flows of material falling down
onto each %
magnetic pole region.
During one period, two episodes of velocity increase towards
the star should be observed. Fig.~\ref{ex-fig2} ({\it left}) illustrates the
temporal behaviour of the different parameters of the He\,{\sc i} $\lambda$10830 profile
depending on the phase of $P_{\rm rot}$ = $42.078^{\rm d}$.
This quasi-resonance line is formed in the high-temperature regions ($T_e\geq15\,000$\,K)
including the innermost disk and the two accretion flows.
The velocity $v_{\rm a}$ (Fig.~\ref{ex-fig2}, {\it top left}) characterizes the kinematics at the disk, and
the velocity  $v_{\rm red}$ the kinematics at the flows.
One can see in Fig.~\ref{ex-fig2} ({\it bottom left}), that $v_{\rm red}$
reaches its maxima  twice per period exactly at the phases where $\left<B_{\rm z}\right>$ has
the positive and negative extrema \citep[see][Fig.\,4]{hubrig2011}.
These are the phases when the magnetic poles cross near the line of sight. The behaviour of
$v_{\rm a}$  is similar, but with a smaller amplitude: the accretion process from the
innermost disk is just beginning. The Pa$\gamma$ line is formed over a more
extended region of the disk, and the accretion flows in this line are less
noticeable. The behaviour of the He\,{\sc i} line parameters fully confirms the MA
model predictions.

Unlike HD\,101412, the line profiles of HD\,104237 do not show absorption
related to the accretion flow (apart from March 25, 2013), see Fig.~\ref{ex-fig1}
({\it right}). However, screening of the stellar limb by the accretion flow appears
as a depression in the red wing of the emission profile: the velocity
of the red wing edge $v_{\rm red}$ and the equivalent width $EW$ {\bf } decrease.
Fig.~\ref{ex-fig2} ({\it right}) shows that variations of these parameters for both
IR lines have a notable correlation. Unlike $v_{\rm red}$, the velocity of
the blue wing edge $v_{\rm blue}$ of these lines does not correlate with $EW$. This
confirms our assumption that the depression of the profile red wing is linked to the
flow passing across the line of sight.

\articlefiguretwo{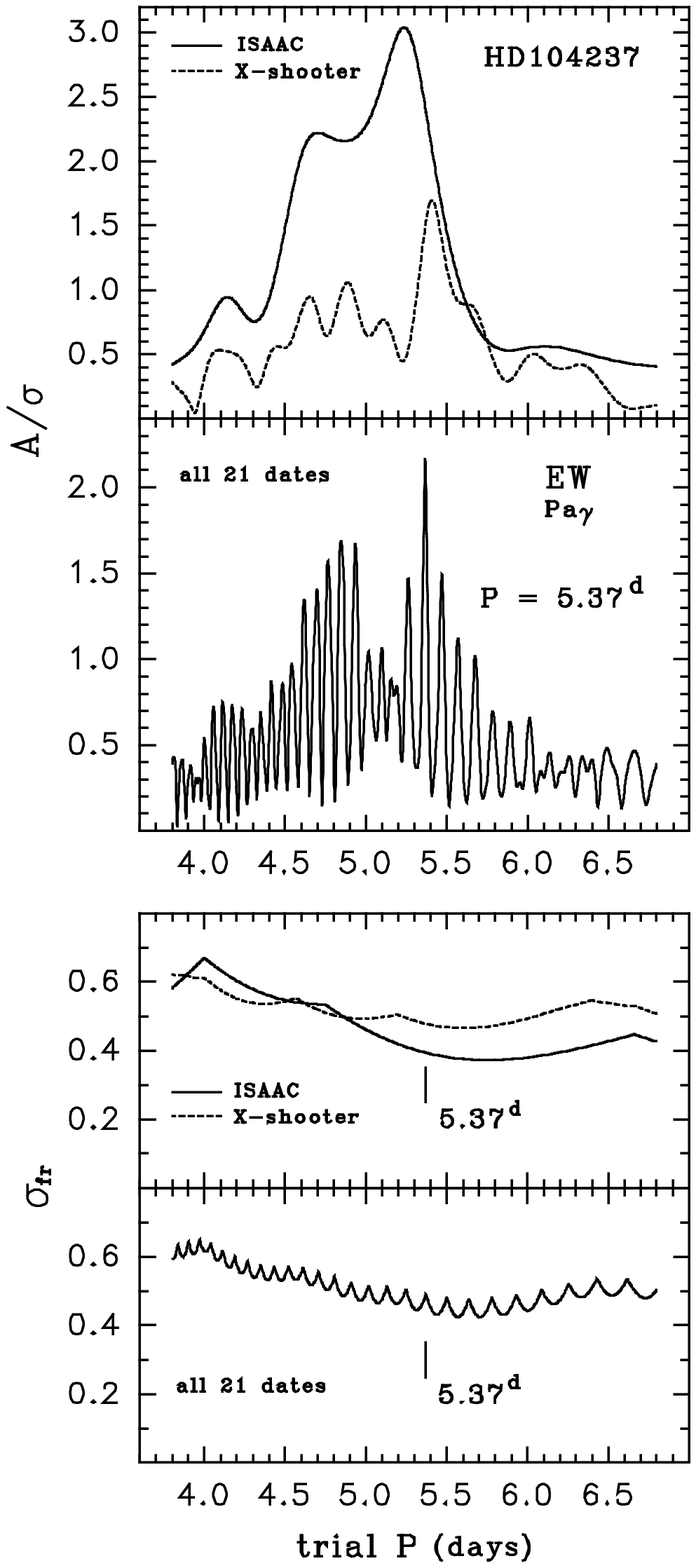}{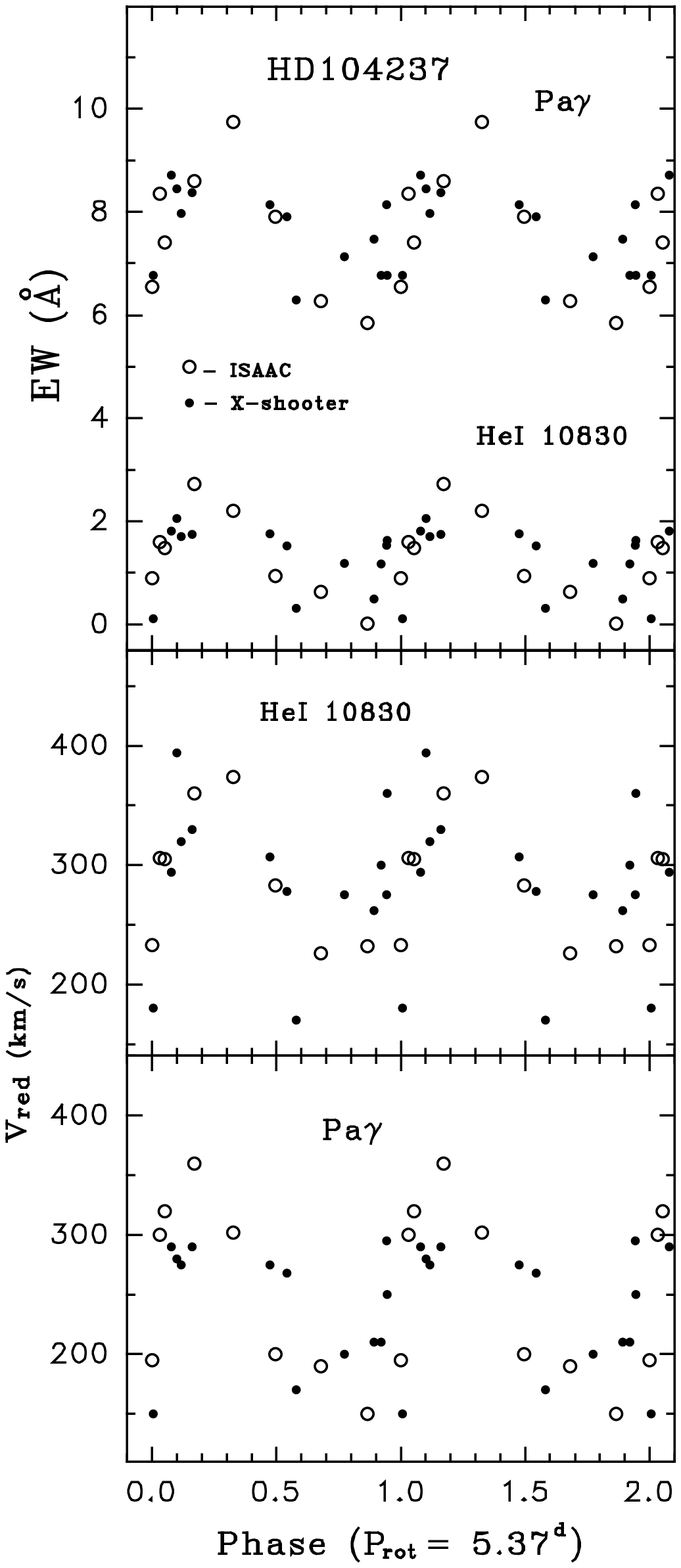}{ex-fig3}{\emph{Left:} Periodograms for
$EW$ variations of the Pa$\gamma$ line in the spectrum of HD\,104237. \emph{Right:}
Phase dependencies of the parameters $EW$ and $v_{\rm red}$ in the spectrum of
HD\,104237. }

We tried to determine the rotation period, $P_{\rm rot}$, of HD\,104237 using a
method based on fitting phase dependencies for each value of the trial period $P$
with a sinusoid for a wide range of $P_{\rm rot}$ (3.8 -- 6.8 days) with a step
width of $0.001^{\rm d}$. The parameters of the sinusoid, such as amplitude $A$,
constant coefficient C, and initial phase $\varphi_{o}$ were determined using a
standard Least-Square method for each value of $P_{\rm rot}$. Fig.~\ref{ex-fig3}
({\it left}) illustrates the periodograms calculated for the parameter $EW$ of the
Pa$\gamma$ line, where $\sigma$ is the standard deviation of values  from the sine
line. We calculated also the periodogram for the parameter $\sigma_{\rm fr}$ which
is the mean square value of the fractional remainder after dividing all time
intervals between neighboring observations by a trial period $P_{\rm rot}$. If the
intervals are close to an integer number of $P_{\rm rot}$, then $\sigma_{\rm fr}$ is
close to zero. In our case we have found $P_{\rm rot}$ = $5.37^{\rm d}$ for the
parameter $EW$ of the Pa$\gamma$ line and have been convinced that this value does
not correspond to a minimum of the parameter $\sigma_{\rm fr}$. Therefore, our
estimation of $P_{\rm rot}$ is not a result of the sampling dates. We plotted
periodograms for parameters $EW$ and $v_{\rm red}$ for the lines He\,{\sc i}
$\lambda$5876, $\lambda$10830 and Pa$\gamma$: for ISAAC data (8 dates), X-shooter
data (13 dates), and all of the dates together (21 dates). The total number of
obtained periodograms was 14, the parameters of the sinusoid $A$ and $\varphi_{o}$
were determined for each case. After averaging, the results were: $P_{\rm rot}$ =
$5.37^{\rm d}\pm0.03^{\rm d}$, $\varphi_{o}$ = $0.030\pm0.041$. The phase
dependencies of the parameters $EW$ and $v_{\rm red}$ for different lines in the
spectrum of HD\,104237 are illustrated in Fig.~\ref{ex-fig3} ({\it right}). Using
standard formulas, we derived the inclination angle $i$. Several estimations of the
stellar radius $R_{*}$ were given in \citet{fumel2012} with a mean value of
$2.95\pm0.11$\,$R_{\odot}$. The following values of $v$\,sin\,$i$ were published in
different papers: $12\pm2$\,km\,s$^{-1}$ \citep{donati1997}, $10\pm1$\,km\,s$^{-1}$
\citep{acke2004}, and $8\pm1$\,km\,s$^{-1}$ \citep{cowley2013}. Adopting
$v$\,sin\,$i$\,=\,$10\pm2$\,km\,s$^{-1}$, we derived $i = 21^\circ\pm4^\circ$.

\articlefiguretwo{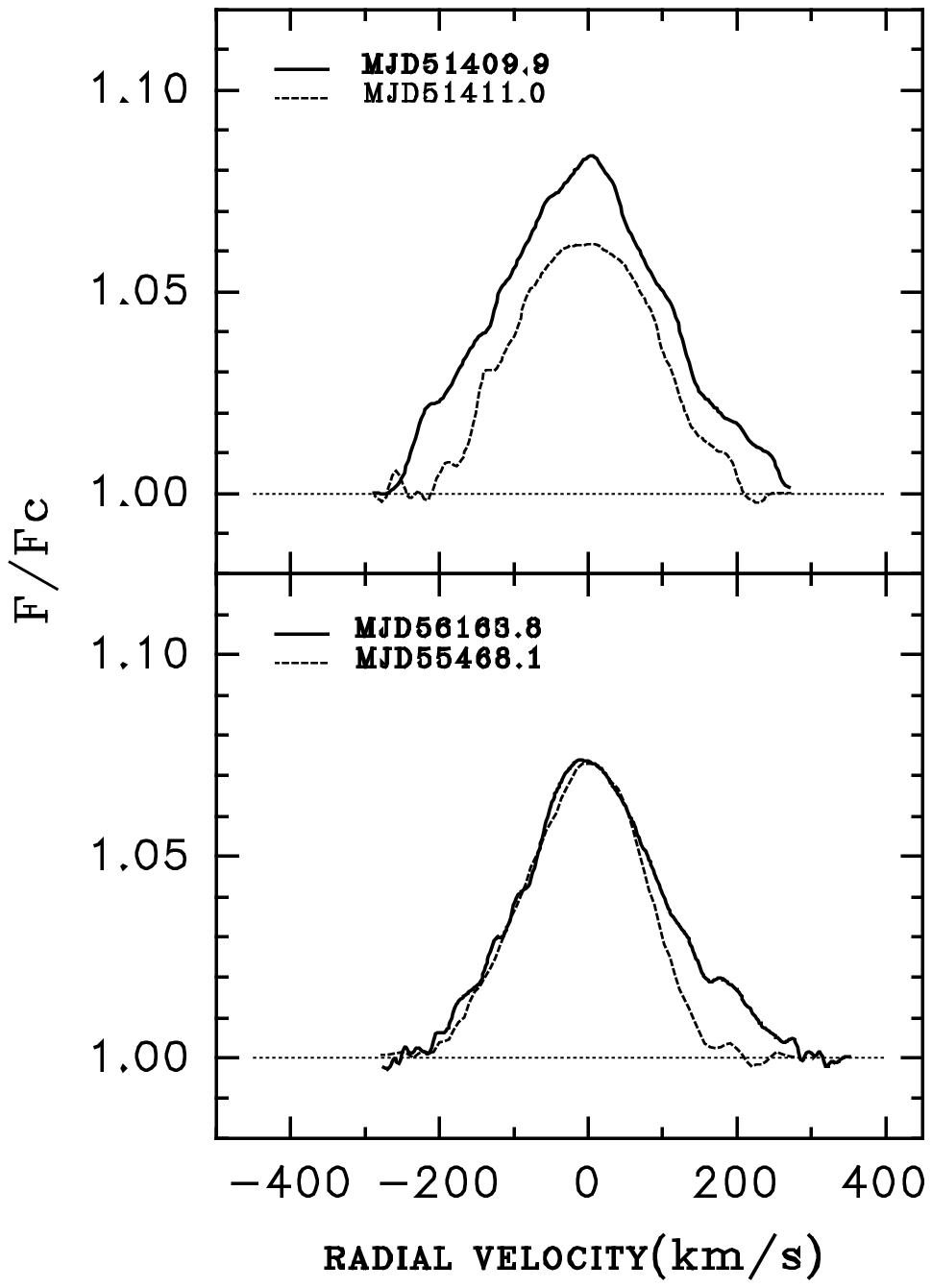}{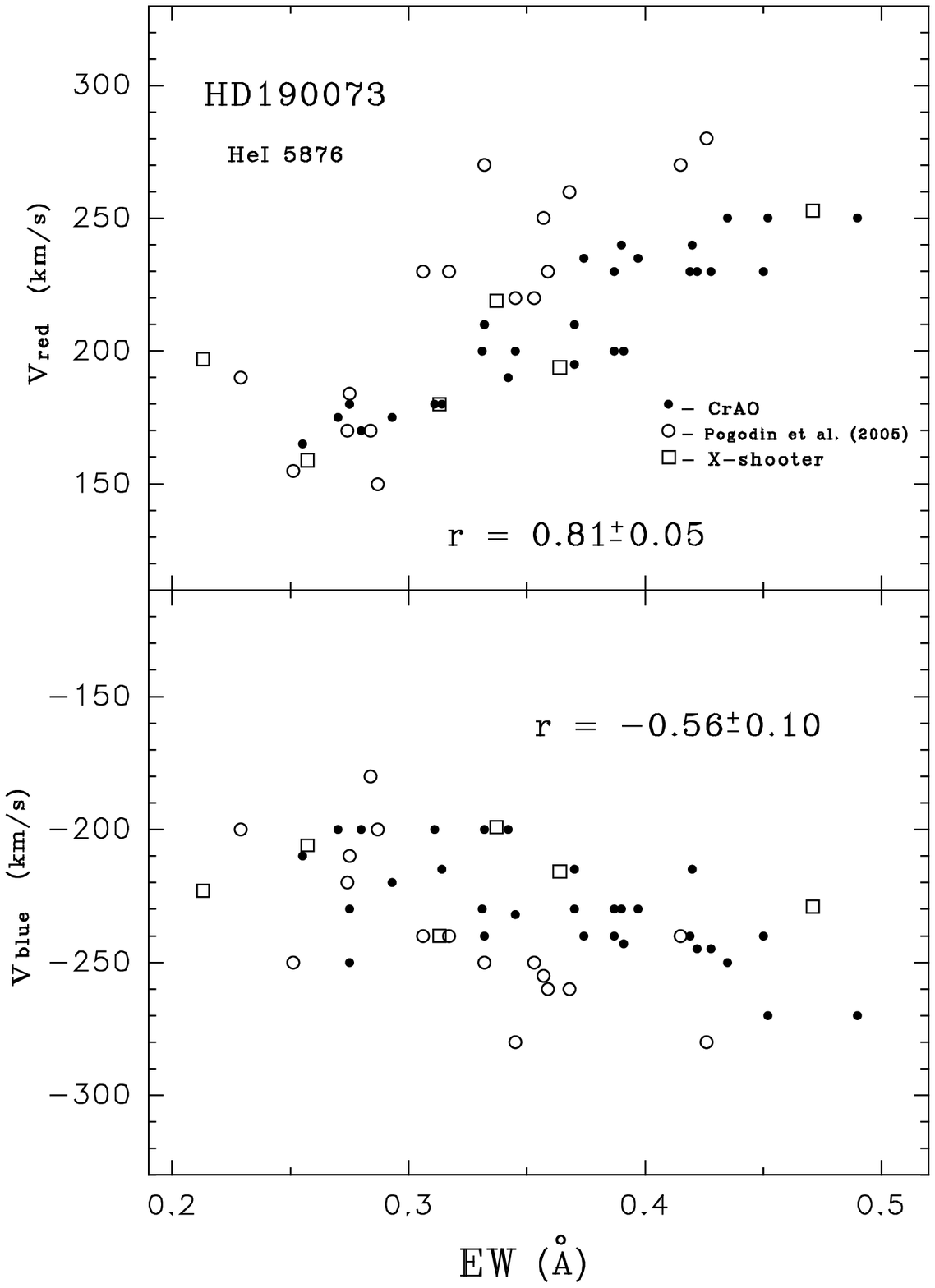}{ex-fig4}{\emph{Left:} Two types of the
He\,{\sc i} $\lambda$5876 line profile in the spectrum of HD\,190073. \emph{Right:}
Correlations of the velocities $v_{\rm red}$ and $v_{\rm blue}$ with $EW$ of the
He\,{\sc i} $\lambda$5876 line profile in the spectrum of HD\,190073.}

For HD\,190073, we had only 11 spectra in the IR region in the period 2010 -- 2013,
which was not enough for a detailed study. The analysis of 51 optical spectra near
the He\,{\sc i} $\lambda$5876 line showed the presence of a large-amplitude
variability on the timescale from 1 day to several years. We separated two main
types of this variability, which is illustrated in Fig.~\ref{ex-fig4} ({\it left}):
{\it a)} a global change of intensity of the line without variations in the shape of
the profile (type~I), and {\it b)} a distortion of the red wing of the emission
profile, as also observed for HD\,104237 (type~II). Fig.~\ref{ex-fig4} ({\it right})
shows the correlations of $v_{\rm red}$ and $v_{\rm blue}$ with the $EW$ of the
line. In the case of the $v_{\rm blue}$/$EW$ dependency, a notable but rather weak
correlation is observed ($r = -0.56\pm0.10$). This correlation corresponds to the
type~I variability. The $v_{\rm red}$/$EW$ correlation is much stronger ($r
=0.81\pm0.05$), it corresponds to both variability types~I and II. We can see that
the contribution of the type~II variability is rather strong in the global behavior
of the He\,{\sc i} $\lambda$5876 line profile variations in the spectra of
HD\,190073. Unfortunately, other attempts to exactly determine the $P_{\rm rot}$ of
HD\,190073 were not successful. However, the behaviour of the $v_{\rm red}$/$EW$
parameter confirms (as in the case of HD\,104237) our assumption of a rotating
accretion flow that distorts the red wing of the emission He\,{\sc i} $\lambda$5876
profile. We assume that the magnetic field configuration of HD\,190073 can be more
complex than a simple dipole, and the whole picture of spectral variability is less
simple.

\articlefiguretwo{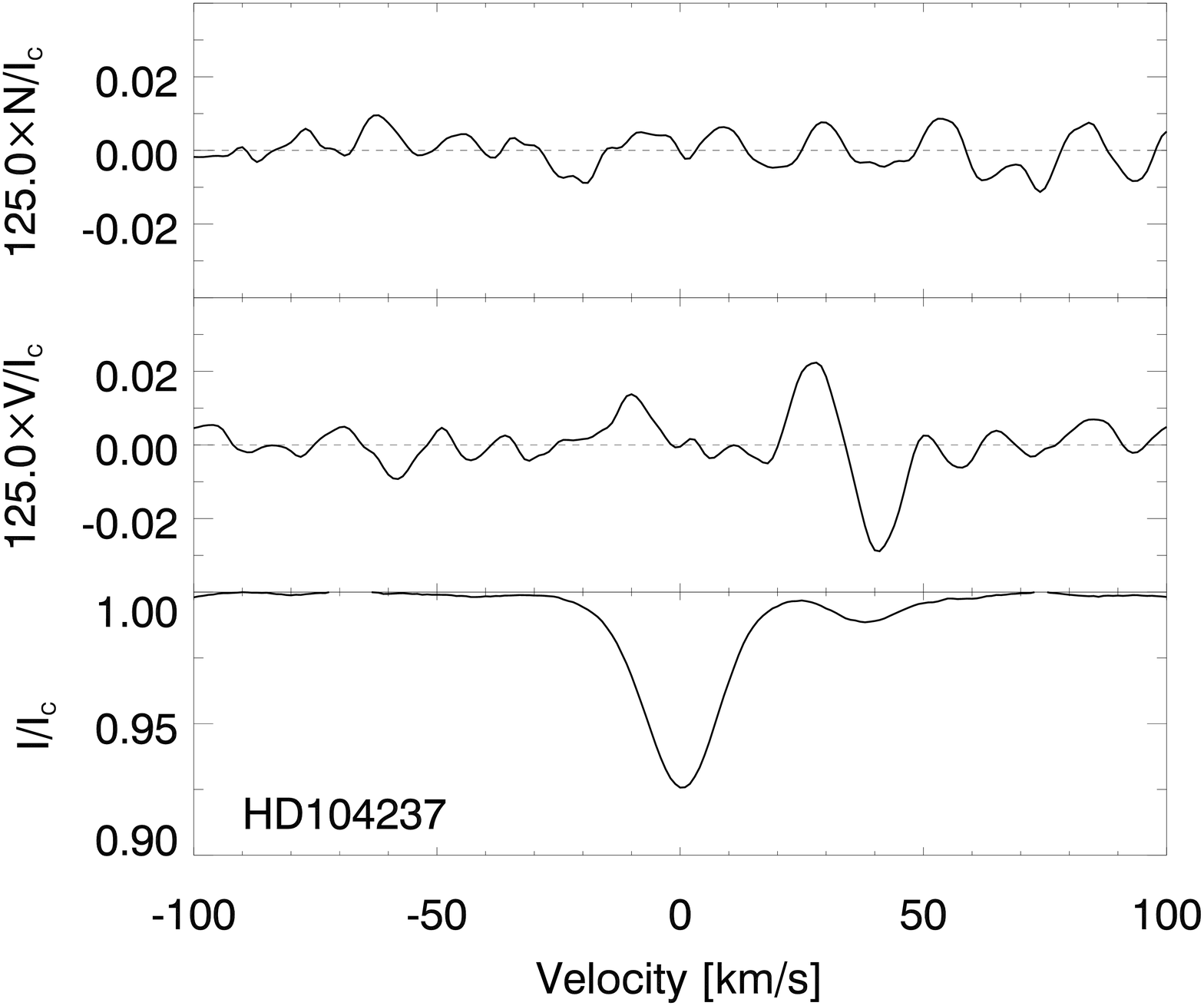}{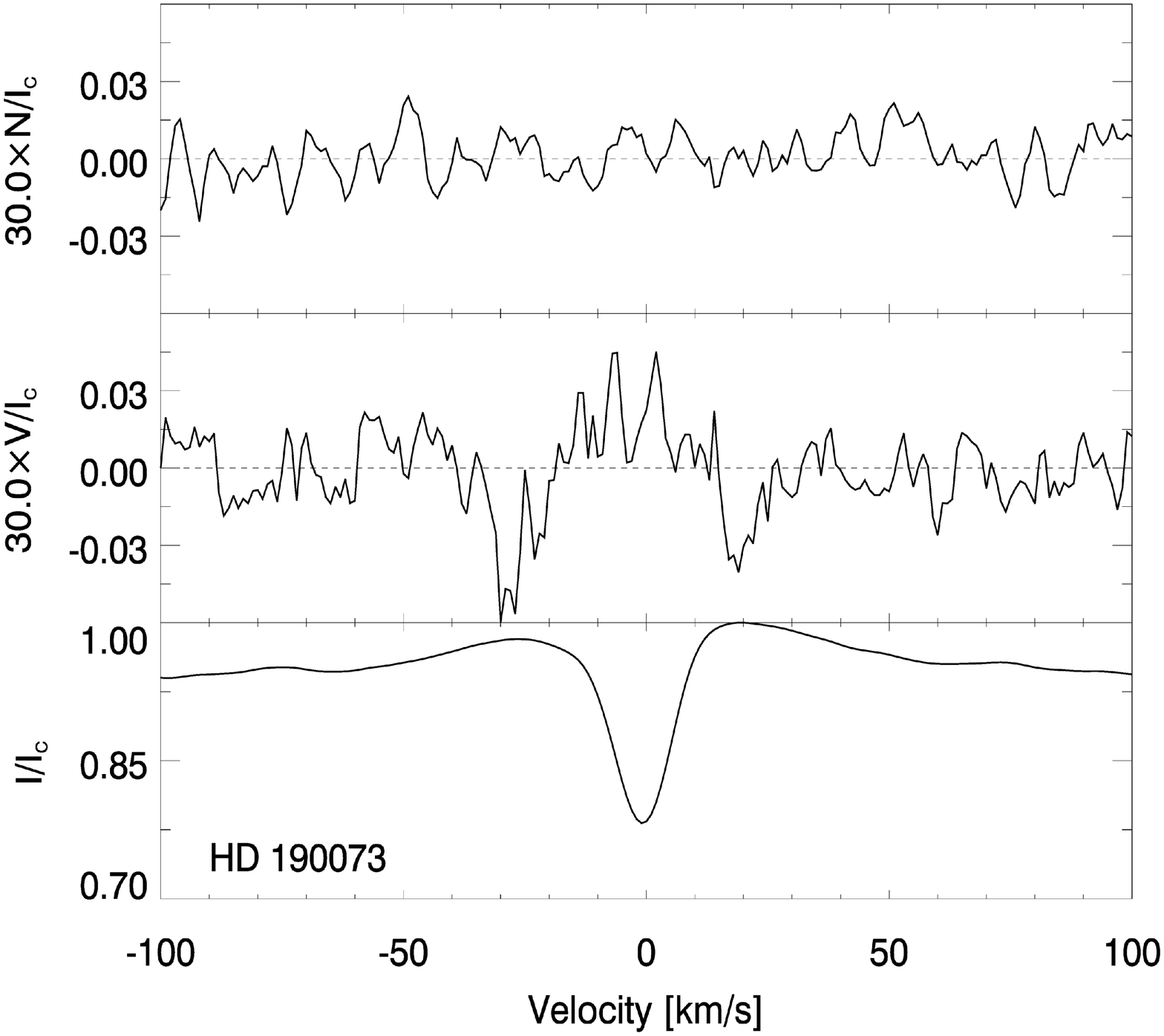}{ex-fig5}{ SVD spectra of HD\,104237 and
HD\,190073 obtained on May 3, 2010 and August 7, 2012, respectively. The upper
spectra represent the diagnostic null spectra, which are associated with the Stokes
$V$ spectra by using subexposures with identical waveplate orientation. Below the
null spectra, we present the Stokes $V$ and Stokes $I$ spectra. The $V$ and null
profiles were expanded by a factor of 125 and 30 for HD\,104237 and HD\,190073,
respectively, and shifted upwards for better visibility. }

Magnetic field measurements were done using high-resolution spectra of
HD\,104237 and HD\,190073 obtained using the HARPS instrument in polarimetric mode,
installed at the 3.6-m
ESO telescope (Chile) and retrieved from the ESO archive. We used the multiline
Singular Value Decomposition (SVD) method for Stokes Profile Reconstruction
\citep{carroll2012}. For HD\,104237,  we can see polarization features corresponding
to the primary component (rather marginal) and to the secondary component of this
binary system (Fig.~\ref{ex-fig5}, {\it left}). The latter is quite notable with
$\left<B_{\rm z}\right>=128\pm10$\,G.
This is the first discovery of the magnetic field of the secondary of HD\,104237.
Another situation is found for HD\,190073 .
Instead of one polarization feature corresponding to the SVD absorption line, we see
two features of opposite signs centered at emission wings around the central
absorption (Fig.~\ref{ex-fig5}, {\it right}).
It is possible that the CS component
has a significant contribution to the general configuration of the magnetic field of
HD\,190073.

\acknowledgements This work was supported by the Basic Research Program of the
Presidium of the Russian Academy of Sciences P-21 ``Non-stationary phenomena in
objects of the Universe''. N.A.D. acknowledges the support of the PCI/MCTI
grant under the project 302350/2013-6 and the St.~Petersburg State University
for research grant 6.38.18.2014.
We would like to thank I.~Ilyin for fruitful discussions on the rotation
periods.

\end{document}